\documentstyle[aps,preprint,tighten,prd,epsf,amsfonts,amsbsy]{revtex}
\begin{document}
\title{Modulating the experimental signature of a stochastic
  gravitational wave background}

\author{Lee Samuel 
Finn$^{(1)}$\cite{bya}, Albert Lazzarini$^{(2)}$\cite{byb}
}
\address{
$^{(1)}$  Center for Gravitational Physics and Geometry, Department of 
Physics, and Department of Astronomy and Astrophysics, The Pennsylvania 
State University, University Park PA 16802\\
$^{(2)}$LIGO Laboratory, California Institute of Technology, 
Pasadena California 91125
}                                     
\draft
\maketitle
\begin{abstract}
  Detecting a stationary, stochastic gravitational wave signal is
  complicated by impossibility of observing the detector noise
  independently of the signal.  One consequence is that we require at
  least two detectors to observe the signal, which will be apparent in
  the cross-correlation of the detector outputs.  A corollary is that
  there remains a systematic error, associated with the possible
  presence of correlated instrumental noise, in any observation aimed
  at estimating or limiting a stochastic gravitational wave signal.
  Here we describe a method of identifying this systematic error by
  varying the orientation of one of the detectors, leading to separate
  and independent modulations of the signal and noise contribution to
  the cross-correlation.  Our method can be applied to measurements of
  a stochastic gravitational wave background by the ALLEGRO/LIGO
  Livingston Observatory detector pair.  We explore --- in the context
  of this detector pair --- how this new measurement technique is
  insensitive to a cross-correlated detector noise component that can
  confound a conventional measurement.
\end{abstract}
\pacs{PACS numbers: 04.80.Nn, 04.80.-y, 95.55.Ym}

\section{Introduction}

While not yet detected, we can be sure that the Earth is bathed in a
stochastic gravitational wave background. Contributions to that
background arise from the confusion limit of large numbers of distant,
conventional gravitational wave sources (e.g., binary systems), early
universe physics (e.g., parametric amplification of fluctuations
during an inflationary phase), and possibly other ``exotic'' physics
(e.g., cosmic strings) \cite{maggiore00a}. One of the goals of the
ground-based gravitational wave detectors now operating or under
construction
\cite{hamilton97a,cerdonio97a,astone97a,luck00a,coles00a,blair00a,marion00a,ando00a}
is to detect or place limits on the amplitude and spectrum of this
stochastic gravitational wave background.

A single, isolated gravitational wave detector cannot distinguish
between instrumental noise and a weak, stationary cosmic gravitational
wave background radiation.  At least two detectors are needed,
in which case the stochastic signal will be apparent in their
cross-correlation.  More particularly, the two detectors must
\emph{i)} have an overlapping frequency response, 
\emph{ii)} have a  separation shorter than the wavelength of their
overlapping response, and 
\emph{iii)} both sample the same polarization state of the
incident radiation.
The LIGO Livingston Observatory (LLO) interferometric detector
\cite{abramovici92a,coles00a} and the Louisiana State University
ALLEGRO cryogenic acoustic detector
\cite{hamilton97a,solomonson94a,mauceli96a}, separated by 42.3~km,
constitute such a detector pair, capable of providing an experimental
bound on the stochastic gravitational wave background at approximately
900~Hz.

Unfortunately, a weak, stationary stochastic signal cannot be
distinguished from a similarly weak, stationary noise background that
is correlated between the two detectors.  Instrumental noise arising
from the environment may lead to such a correlated detector noise;
consequently, a convincing case must be made that no terrestrial noise
source is responsible for any observed correlation.  This is a
daunting experimental challenge. Any technique that can improve
our ability to discriminate between a stochastic gravitational wave
signal and a weak correlation of terrestrial origin should be pursued.

Since the signal contribution to the cross-correlation depends on the
relative orientation of the detectors (which determines their
sensitivity to the two different polarization states), changing the
orientation of one of the detectors will modulate the signal
contribution to the cross-correlation in a predictable way, allowing
us to distinguish the signal from the correlated noise and leading to
a significantly improved estimate or bound on the in-band amplitude of
a stochastic gravitational wave background. The technique of
introducing a controlled signal modulation in order to identify and
eliminate systematic environmental effects was conceived by Dicke
\cite{dicke46a} as the switching radiometer during the development of
radar.

The ALLEGRO group has used a planned relocation of their cryogenic
detector to a new laboratory in order to implement the capability to
re-orient the detector between data-taking periods\cite{allegroII}.
This capability now allows for a modulation of the gravitational wave
contribution to the detector noise cross-correlation in exactly the
manner described. Here we discuss the details of the modulation and
how it can be used to improve the reliability of the estimate or limit
that we can place on the amplitude of a stochastic gravitational wave
background near 900 Hz.

The basic scheme for detecting a stochastic gravitational wave
background using two or more gravitational wave detectors was
described in \cite{michelson87a} and further elucidated in
\cite{christensen92a,flanagan93a,allen99b}.  In Section
\ref{section:aa} we review that work and extend it to the case where
the two detectors share a cross-correlated noise component.  In
Section \ref{section:bb} we apply these results to the specific case
of the ALLEGRO/LLO detector pair, providing the relevant geodetic and
physical parameters that are needed to quantify the magnitude of the
correlation and describing how to extract the correlation from
observations.  In Section \ref{section:cc} we present results of
numerical calculations based on the initial LIGO instrumentation and its
proposed upgrade. 

\section{The cross-correlated detector output}
\label{section:aa}

\subsection{The cross-correlation statistic}

The output of each detector --- ALLEGRO or LLO --- is a single time
series, which is the sum of instrumental noise and a projection of the
incident gravitational wave strain.  Denote the output of LLO as $s_L$
and the output of ALLEGRO as $s_A$ and define the correlation of $s_L$
and $s_A$ over an integration time $T_{\text{int}}$ by
\cite{michelson87a,christensen92a,flanagan93a,allen99b}
\begin{mathletters}
\begin{eqnarray}
    C(\bbox{\Omega_A},\bbox{\Omega_L}) &:=& \left<s_{A},S_{L}\right>\\
    &:=& \int_{-T_{\text{int}}/2}^{T_{\text{int}}/2}dt
    \int_{-T_{\text{int}}/2}^{T_{\text{int}}/2}dt'\,
    s_A(t)s_L(t')Q(t-t';\bbox{\Omega_A},\bbox{\Omega_L}),
    \label{eqn:C-defn}
\end{eqnarray}
where 
\begin{eqnarray}
    \bbox{\Omega}_{A} &:=& \text{Angles describing the orientation of 
    ALLEGRO}\\
    \bbox{\Omega}_{L} &:=& \text{Angles describing the orientation of 
    LLO}.
\end{eqnarray}
\end{mathletters}
We discuss the choice of integration kernel
$Q\left(\tau;\bbox{\Omega}_{A},\bbox{\Omega}_{L}\right)$ below.  

To evaluate the expectation value of $C$ in the presence of signal and
noise, write
\begin{equation}
    s_{*} := n_{*} + h_{*},
\end{equation}
where $n_{*}$ is the noise and $h_{*}$ is the signal in detector $*$. 
The signal is, of course, statistically independent of the noise; 
consequently, 
\begin{mathletters}
\begin{eqnarray}
    \mu &:=& \overline{C} \\
    &:=& \overline{\left<h_{A},h_{L}\right>} + 
    \overline{\left<n_{A},n_{L}\right>},
\end{eqnarray}
where the overbar represents an ensemble average. Introducing also 
the variance of $C$,
\begin{equation}
    \sigma^{2} := \overline{C^{2}}-\overline{C}^{2},
\end{equation}
\end{mathletters}
we define the dimensionless signal-to-noise ratio (SNR)
\begin{equation}
    \rho := {\mu/\sigma}. \label{eqn:defn-rho}
\end{equation}

As described above, the integration kernel in equation \ref{eqn:C-defn}
is at our disposal. If the detector and signal noise spectral
densities are known, then $Q$ can be chosen to maximize the signal to
noise ratio.  Previous work \cite{christensen92a,flanagan93a,allen99b}
has always assumed that the contribution of the noise
cross-correlation (e.g., $\overline{\left<n_{A},n_{L}\right>}$) to the
mean $\mu$ vanishes.  In Appendix \ref{app:Q} we consider the more
general case where the noise contribution to the ensemble mean
cross-correlation is non-zero. In this case the kernel $Q$ that
maximizes the SNR can be conveniently expressed in the frequency
domain as
\begin{mathletters}
\begin{equation}
\widetilde{Q}(f; \bbox{\Omega_A},\bbox{\Omega_L}) := 
{\gamma(f;\bbox{\Omega_A},\bbox{\Omega_L})\Omega_{\text{{GW}},0}(f)\over
f^3 \biggl( S_A(f)S_L(f) + S_{AL}(f)^{2} \biggr)} ~,
\label{eqn:optimal}
\end{equation}
where 
\begin{eqnarray}
    S_{A} &:=& \left(\text{ALLEGRO noise power spectral density}\right)\\
    S_{L} &:=& \left(\text{LLO noise power spectral density}\right)\\
    S_{AL} &:=& \left(\text{ALLEGRO/LLO noise cross spectral 
    density}\right)\\
    \Omega_{\text{{GW}},0} &:=& \left(\begin{array}{l}
    \text{expected stochastic signal spectrum expressed as a }\\
    \text{fraction of the closure density in logarithmic frequency}
    \end{array}\right) \label{eqn:Omega0}
\end{eqnarray}
\end{mathletters}
and $\gamma(f;\bbox{\Omega_A},\bbox{\Omega_L})$ is the {\em overlap
reduction function,} which describes the amplitude of the correlation
of the gravitational wave signal between the two detectors as a
function of their relative orientation
\cite{michelson87a,christensen92a,flanagan93a}.

\subsection{The overlap reduction function}

The overlap reduction function reflects frequency-dependent
correlation of the gravitational wave signal in the two detectors and
is expressed in terms of their relative orientation and separation.
To express $\gamma$ for the ALLEGRO and LLO detectors, focus first on
the two arms of the LLO detector, which define a plane (see figure
\ref{figure:coords}).  Let $\widehat{\bbox{n}}_{x}$ be the unit vector
along the projection onto this plane of the unit vector pointing from
the LLO vertex toward the ALLEGRO bar's midpoint.  Taking
$\widehat{\bbox{n}}_{z}$ to be orthogonal to the plane and in the
direction of increasing altitude, define $\widehat{\bbox{n}}_{y}$ in
the plane to form, with $\widehat{\bbox{n}}_{x}$ and
$\widehat{\bbox{n}}_{y}$ a right-handed coordinate system.

By construction the arms of the LLO detector also lie in this plane
and, by convention, they are referred to as the $X$ and $Y$ arms.
Denoting the unit vector in the direction from LLO vertex along the
$X$ arm as $\widehat{\bbox{X}}$ and the unit vector in the direction
from the vertex along the $Y$ arm as $\widehat{\bbox{Y}}$,
$\widehat{\bbox{X}}$ and $\widehat{\bbox{Y}}$ form a right-handed
coordinate system with $\widehat{\bbox{n}}_{z}$.  It is convenient to
introduce the bisector of the pair
$(\widehat{\bbox{X}},\widehat{\bbox{Y}})$, in the direction of
increasing $\widehat{\bbox{X}}$ and $\widehat{\bbox{Y}}$, and define
the angle $\sigma_{L}$ to be the angle between the bisector and
$\bbox{n}_{x}$. Then, writing the $TT$-gauge gravitational wave strain
as $h_{ij}$, the LLO detector responds to the superposition
\begin{equation}
    h_{L} := h_{ij}d_{L}^{ij}
\end{equation}
where
\begin{equation}
    \bbox{d}_L(\sigma_{L}) := 
    {1\over2}\left[
    \sin(2\sigma_L)\left(
    \hat{\bbox{n}}_x\otimes\hat{\bbox{n}}_x -
    \hat{\bbox{n}}_y\otimes\hat{\bbox{n}}_y
    \right) - 
    \cos(2\sigma_L)\left(
    \hat{\bbox{n}}_x\otimes\hat{\bbox{n}}_y +
    \hat{\bbox{n}}_y\otimes\hat{\bbox{n}}_x
    \right)
    \right]
    \label{eqn:dL}
\end{equation}
of the incident gravitational waves. In computing the overlap 
reduction function, $\bbox{d}_{L}$ characterizes the LLO detector.

Assume that symmetry axis of the ALLEGRO bar lies in the
$\widehat{\bbox{n}}_{x}\times\widehat{\bbox{n}}_{y}$ 
plane.\footnote{This is a
good approximation for the ALLEGRO/LLO detector pair.} and define
\begin{equation}
    \sigma_{A} := 
    \left(\text{the angle between the ALLEGRO symmetry axis and
    $\widehat{\bbox{n}}_{x}$}
    \right).
\label{eqn:sigmaA}
\end{equation}
In terms of $\sigma_{A}$ the ALLEGRO detector responds to the
superposition
\begin{equation}
    h_{A} := h_{ij}d_{A}^{ij},
\end{equation}
where
\begin{eqnarray}
\bbox{d}_A(\sigma_A) &:=& 
\left(
  \hat{\bbox{n}}_x \cos\sigma_A +
  \hat{\bbox{n}}_y \sin\sigma_A
\right) \nonumber \\ &&\otimes
\left(
  \hat{\bbox{n}}_x \cos\sigma_A +
  \hat{\bbox{n}}_y \sin\sigma_A
\right) -  {1\over3}\bbox{I},
\label{eqn:dA}
\end{eqnarray}
of the incident gravitational waves. In computing the overlap 
reduction function, $\bbox{d}_{A}$ characterizes the ALLEGRO detector.

Finally, define the functions
$\rho_k(\alpha)$ by 
\begin{equation}
\left(\begin{array}{c}
\rho_1(\alpha)\\
\rho_2(\alpha)\\
\rho_3(\alpha)
\end{array}\right)
:= 
{1\over\alpha^2}
\left(\begin{array}{ccc}
5\alpha^2&-10\alpha&5\\
-10\alpha^2&40\alpha&-50\\
5\alpha^2/2&-25\alpha&175/2
\end{array}
\right)
\left(\begin{array}{c}
j_0(\alpha)\\
j_1(\alpha)\\
j_2(\alpha)
\end{array}\right),
\end{equation}
where the $j_k$ are the spherical Bessel functions of order $k$, 
\begin{equation}   
\alpha := {2\pi f L/c}, \label{eqn:alpha}
\end{equation}
and $L$ is the length of the baseline between the two detectors. The 
functions $\rho_{k}(\alpha)$ characterize the frequency dependent 
part of the sensitivity of the detector pair to a stochastic  
gravitational wave background. 

Taking all these pieces together, the overlap reduction can be 
expressed 
\begin{eqnarray}
\label{eqn:ba}
\gamma(f;\bbox{\Omega_A},\bbox{\Omega_L}) &:=& 
\rho_1(\alpha) \bbox{d}_A:\bbox{d}_L +
\rho_2(\alpha) \left(\bbox{\hat n}_{x}\cdot\bbox{d}_A\right)\cdot
\left(\bbox{d}_{L}\cdot\bbox{\hat n}_{x}\right)\nonumber\\
&&\qquad{}+
\rho_3(\alpha) \left(\bbox{\hat n}_{x}\cdot\bbox{d}_A\cdot\bbox{\hat 
n}_{x}\right)
\left(\bbox{\hat n}_{x}\cdot\bbox{d}_L\cdot\bbox{\hat n}_{x}\right),
\end{eqnarray}
Table \ref{tbl:coordsL} gives the relevant geographic parameters
describing LLO detectors \cite{althouse01a}; table \ref{tbl:coordsA}
does the same for the ALLEGRO detector\cite{allegroII}.

\section{Application: ALLEGRO and LLO}
\label{section:bb}

\begin{figure}
\begin{center}
\epsfxsize=\columnwidth
\epsffile{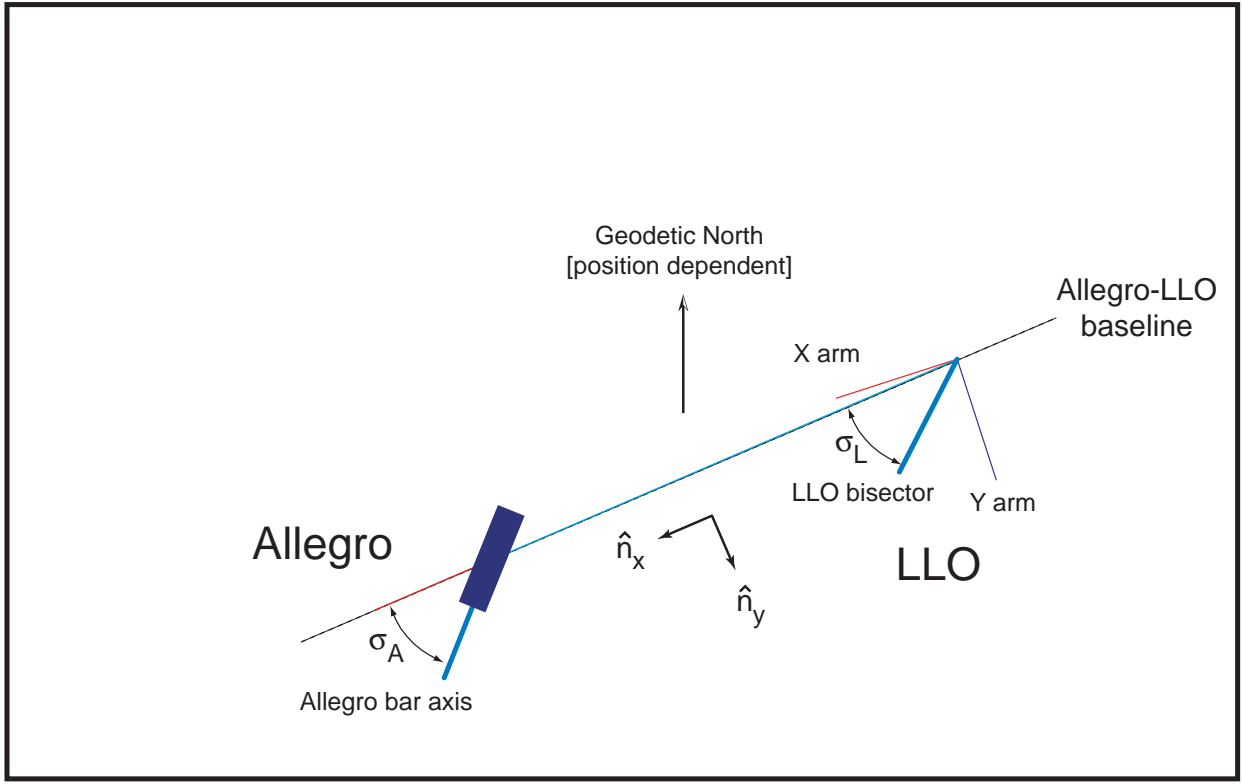}
\end{center}
\caption{A schematic diagram showing how we characterize ALLEGRO and LLO
  orientations with respect to geodetic north and the LLO-to-ALLEGRO
  baseline.}
\label{figure:coords}
\end{figure}

The LLO detector orientation is fixed; however, the orientation of the
ALLEGRO detector may be changed by rotating ALLEGRO in its horizontal
plane (cf.\ figure \ref{figure:coords}).  This degree of freedom is
described by the angle $\sigma_A$, defined in equation
\ref{eqn:sigmaA}.  As $\sigma_A$ varies,
$C(\bbox{\Omega_A},\bbox{\Omega_L})$ will change through the
dependence of $\gamma$ on $\sigma_A$.  To express that variation write
the output of detector $k$ as the sum of a gravitational wave signal
$h_k$ and detector noise $n_k$:
\begin{equation}
s_k(t) := h_k(t) + n_k(t). 
\end{equation}
We can then write the ensemble average of the
correlation $C(\bbox{\Omega_A},\bbox{\Omega_L})$ as a
function of $\sigma_A$:
\begin{mathletters}
\begin{equation}
\overline{C}(\sigma_A) := 
\left(
\overline{<h_A,h_L>} + \overline{<n_A,n_L>}
\right),
\end{equation}
where
\begin{eqnarray}
\overline{\left< h_A,h_L \right>} &:=& T_{\text{int}}\int df\,
{3H_0^2\over20\pi^2}
{\gamma^2(f;\sigma_A)\Omega_{\text{{GW}}}(f)\Omega_{0,GW}(f)
\over f^6 \left( S_A(f)S_L(f)+S_{AL}(f)^{2} \right) },\label{eqn:hh}\\
\overline{\left<n_A,n_L\right>} &:=&
T_{\text{int}}\int df\,
S_{AL}(f;\sigma_A)
{\gamma(f;\sigma_A)\Omega_{\text{{GW}},0}(f)\over 
f^3 \left( S_A(f)S_L(f)+S_{AL}(f)^{2} \right)},\label{eqn:nn}\\
\Omega_{\text{{GW}}} &:=&
\left(
\begin{array}{l}
\text{The actual stochastic gravitational wave spectrum}
\end{array}
\right)
\end{eqnarray}
and
\begin{equation}
S_{AL}(f;\sigma_A) := \left(
\begin{array}{l}
\text{The ALLEGRO/LLO noise cross-spectral density,}\\
\text{which in general may depend on the orientation angle $\sigma_A$}
\end{array}
\right).
\end{equation}
\end{mathletters}
Since $\gamma$ depends on the orientation $\sigma_{A}$ of the ALLEGRO
detector, changing ALLEGRO's orientation changes $\overline{C}$ and
allows us to modulate the gravitational wave contribution to $C$ in a
predictable way. 

\begin{figure}
\begin{center}
\epsfxsize=\columnwidth
\epsffile{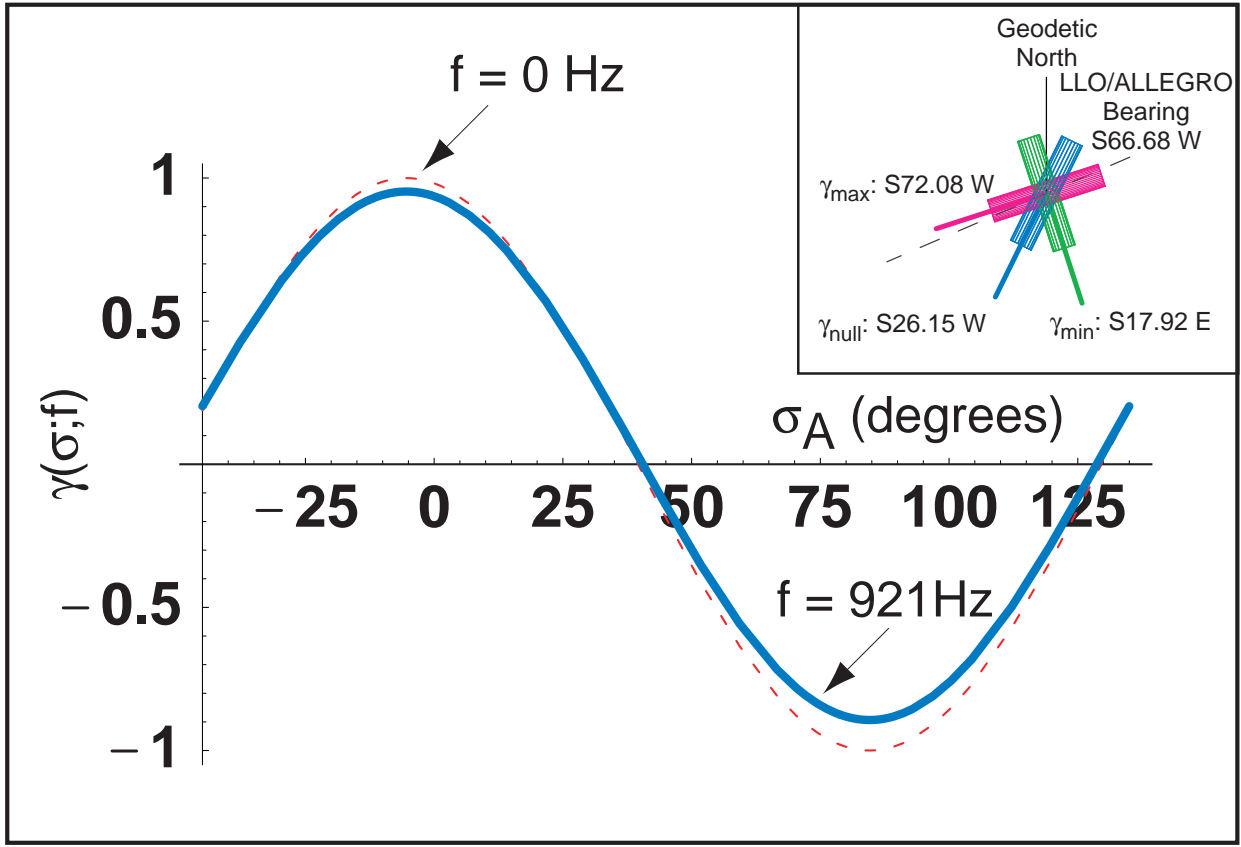}
\end{center}
\caption{The overlap reduction function $\gamma$, which characterizes the
  dependence of the ALLEGRO-LLO correlation function to an isotropic
  stochastic signal, depends on the relative orientation of the two
  detectors. Here we show how this function varies with $\sigma_A$,
  the angle between the ALLEGRO bar axis and the LLO/ALLEGRO baseline
  (cf.\ figure \ref{figure:coords}). The bold, solid line shows the
  variation of $\gamma$ with $\sigma_A$ at the operating frequency of
  the ALLEGRO detector; for comparison, the dashed line shows the same
  quantity at DC. Inset B shows the orientation of the ALLEGRO
  relative to the LLO/ALLEGRO baseline, when $\gamma$ vanishes
  ($\gamma_{\text{null}}$) and takes on its minimum
  ($\gamma_{\text{min}}$) and maximum ($\gamma_{\text{max}}$)
  values.}\label{figure:gamma}
\end{figure}

Signal correlations between LLO and ALLEGRO occur only in narrow bands
centered on the two ALLEGRO bar resonances ($f_{<}$, $f_{>}$; cf.\ 
table \ref{tbl:opChar}).  Over the band bounded by the resonances the
LLO noise power spectral density $S_{L}$ should be approximately
constant and we expect that $\Omega_{\text{GW}}$ will be constant as well.
Additionally, for the ALLEGRO/LLO detector pair $\alpha$ (cf.\ Eq.
\ref{eqn:alpha}) is small (approximately 0.8) and does not change
significantly between the two resonances, so that the overlap
reduction function can be treated as frequency independent where the
integrands in either of equations \ref{eqn:nn} and \ref{eqn:hh} are
significant.  Further assuming that $S_{AL}$, the ALLEGRO-LLO
instrumental noise cross-spectral noise density, is independent of
$\sigma_A$ and much smaller than either $S_{L}$ or $S_{A}$, we obtain
\begin{mathletters}
\begin{equation}
    C(\sigma_{A}) \simeq 
    T_{\text{int}}\Delta f
    {\Omega_{\text{{GW}},0}(f_{0})
      \over f_{0}^{3}S_{A}(f_{0})S_{L}(f_0)}
    \left[\gamma^{2}(f_0;\sigma_{A})
    {3H_{0}^{2}\over20\pi^{2}}
    {\Omega_{\text{GW}}(f_{0})\over
    f_{0}^{3}}
    + \gamma(f_0;\sigma_{A})S_{AL}(f_{0})
    \right],
\label{eqn:C-approx}
\end{equation}
where
\begin{eqnarray}
    \Delta f &:=& 
    \left(
    {\Delta f_{>}\over S_{A}(f_{>})} + 
    {\Delta f_{<}\over S_{A}(f_{<})}
    \right)S_{A}\\
    S_{A} &:=& {1\over2}\left[S_{A}(f_{>}) + S_{A}(f_{<})\right].
  \end{eqnarray}
\end{mathletters}
As $\sigma_{A}$ varies the contribution of the stochastic signal to
$C$ ($\left<h_{A},h_{L}\right>$, which is quadratic in $\gamma$)
varies differently than the contribution of the instrumental noise
($\left<n_{A},n_{L}\right>$, which is linear in $\gamma$).  Figure
\ref{figure:gamma} shows, as a solid line, the dependence of $\gamma$ on
$\sigma_{A}$.  For reference, the dotted line shows the dependence of
$\gamma$ on $\sigma_{A}$ at zero frequency.  Note how $\gamma$ is
approximately sinusoidal in $2\sigma$.  Figure \ref{figure:coords}B
shows a schematic of the ALLEGRO bar orientation corresponding to the
extrema and null of $\gamma$ as shown in figure \ref{figure:gamma}.

Since the two additive terms in equation \ref{eqn:C-approx} depend on
$\gamma(f_0,\sigma_A)$ differently, varying $\sigma_A$ modulates the
contribution to $C$ of any correlated noise differently than it
modulates the contribution of a real signal. We can use this
differential modulation to eliminate the contribution of any
correlated noise $S_{AL}$ that is independent of $\sigma_A$. 
Denote the angle $\sigma_{A}$ for which $\gamma$ is maximized as
$\sigma_{A,\max}$; similarly, denote the angle $\sigma_{A}$ for which
$\gamma$ is minimized as $\sigma_{A,\min}$.  Suppose we make an
observation of duration $T_{\text{int},\max}$ with ALLEGRO oriented at
angle $\sigma_{A,\max}$, and another observation of duration
$T_{\text{int},\min}$ at angle $\sigma_{A,\min}$.  The expectation
value of $C$ for these two observations is
\begin{mathletters}
\begin{equation}
    \label{eqn:yy}
    \overline{C}(\sigma_{\text{max}}) \simeq 
    T_{\text{int},\max} \left(\Delta f
    {\Omega_{\text{{GW}},0}
      \over f_{0}^{3}S_{A}S_{L}}\right)
    \left[
    \left({3H_{0}^{2}\over20\pi^{2}f_{0}^{3}}\right)
    {\gamma_{\text{max}}^{2}\Omega_{\text{GW}}}
    + \gamma_{\text{max}}S_{AL}
    \right]
\end{equation}
and
\begin{equation}
    \label{eqn:xx}
   \overline{C}(\sigma_{\text{min}}) \simeq 
    T_{\text{int},\min} \left(\Delta f
    {\Omega_{\text{{GW}},0}
      \over f_{0}^{3}S_{A}S_{L}}\right)
    \left[
    \left({3H_{0}^{2}\over20\pi^{2}f_{0}^{3}}\right)
    {\gamma_{\text{min}}^{2}\Omega_{\text{GW}}}
    + \gamma_{\text{min}}S_{AL}
    \right].
\end{equation}
where
\begin{eqnarray}
    \gamma_{\max} &:=& \gamma(\sigma_{A,\max})\\
    \gamma_{\min} &:=& \gamma(\sigma_{A,\min})
\end{eqnarray}
\end{mathletters}
The combination of these two observations
\begin{equation}
    C_{0} := {\gamma_{\max}T_{\text{int},\max}{C}(\sigma_{\min})
    -\gamma_{\min}T_{\text{int},\min}{C}(\sigma_{\max})\over
    \gamma_{\max}T_{\text{int},\max}+\left|
    \gamma_{\min}\right|T_{\text{int},\min}}
\label{eqn:C0-defn}
\end{equation}
thus has an expectation value that is independent of the correlated
noise $S_{AL}$. In the present circumstance, 
\begin{equation}
\gamma_{0} := \gamma_{\text{max}} \simeq -\gamma_{\text{min}}
\end{equation}
If we also make the observations of equal duration,
\begin{equation}
    T_{\text{int},\max} = T_{\text{int},\min} = T_{\text{int}}/2
\end{equation}
then, following past convention and defining the signal-to-noise ratio
$\rho_{0}$ of the observation $C_{0}$ as the ratio of $C_{0}$ to its
ensemble variance we have\footnote{Here we assume that both $S_{A}$
  and $S_{L}$ are much greater than either $|S_{AL}|$ or the
  corresponding power spectral density of the stochastic signal
  $S_{h}$. Were this not the case we would likely be able to identify
  the origin of the correlated noise and either isolate the detector
  pair from it, or regress it from the data during analysis.}
\begin{equation}
    \label{eqn:zz}
    \overline{\rho_{0}} 
    \approx {3H_{0}^{2}\over10\pi^{2}} \sqrt{T_{\mbox{int}}\Delta f }
    {\gamma_{0}\Omega_{\text{GW}}(|f_{0}|)\over 
    f_{0}^{3}\sqrt{S_{A}(f_{0})S_{L}(f_{0})+S_{AL}(f_{0})^{2}}}.
\end{equation}

\section {Numerical Results}
\label{section:cc}

Table \ref{tbl:opChar} describes the theoretical limiting operating
characteristics of the current ALLEGRO
\cite{hamilton97a,solomonson94a,mauceli96a}, the initial LIGO
Livingston detector \cite{LIGOE95001802}, and the planned upgrades to
ALLEGRO \cite{allegroII} and LIGO \cite{ligoII}.  The last two rows
give the theoretical limiting sensitivities (90\% confidence bounds)
for $\Omega_{\text{GW}}(900 Hz)$ that are achievable by the detector
pairs \emph{LIGO I $+$ ALLEGRO} and \emph{Advanced LIGO $+$ Upgraded
  ALLEGRO} in a one year observation.  These limits can be reached
only by identifying and accounting for non-gravitational wave
inter-detector correlations.  Unaccounted correlations of
non-gravitational wave origin introduce a systematic error that
quickly becomes the limiting factor in an upper limit determination.

\begin{figure}
\begin{center}
\epsfxsize=\columnwidth
\epsffile{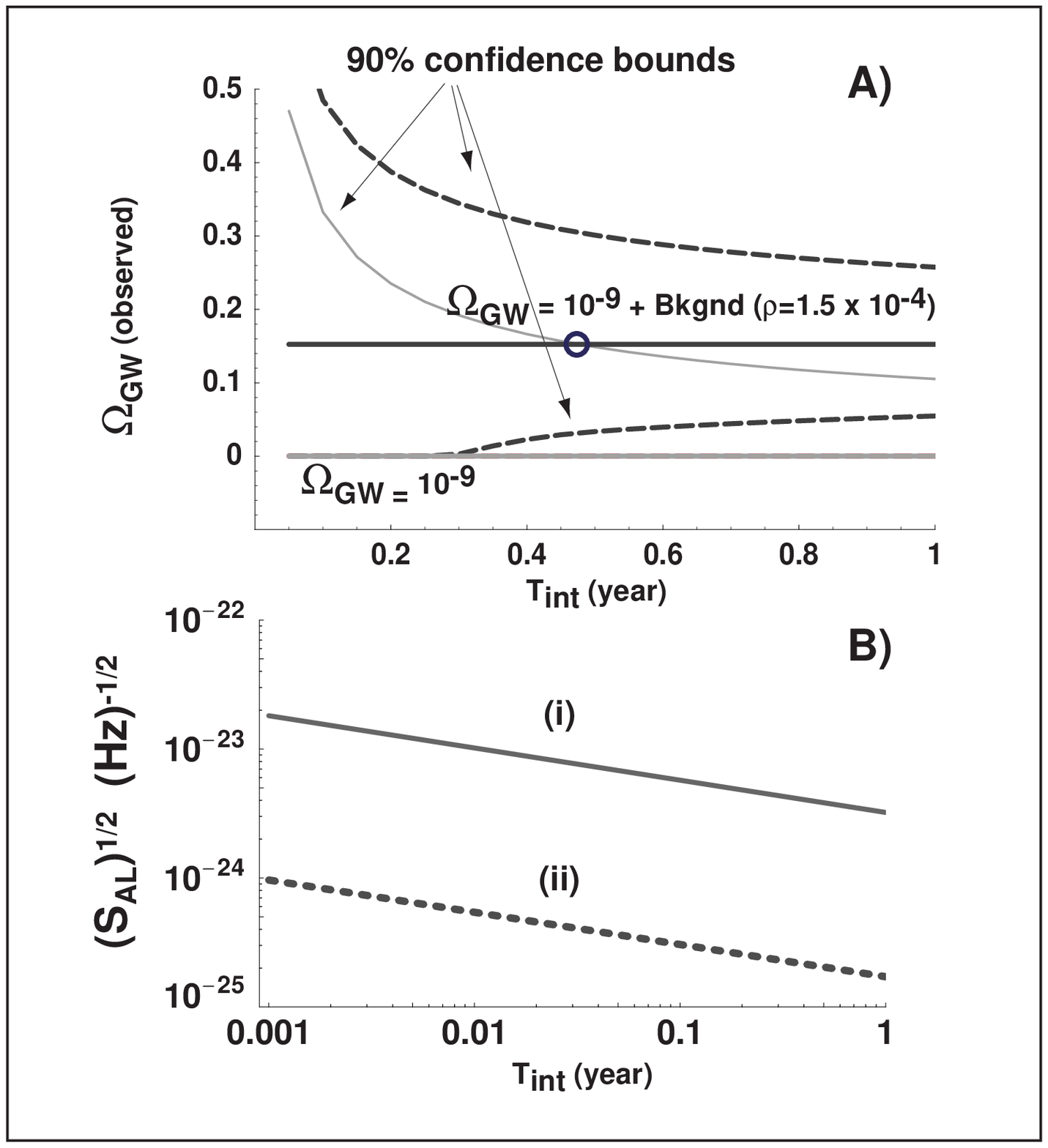}
\end{center}
\caption{
  A) The dashed lines mark the expected 90\% confidence interval, as a
  function of the observing time, on a stochastic gravitational wave
  background when a much larger, but unaccounted for, correlated
  terrestrial noise source is present with a cross spectral density
  amplitude just $10^{-4}$ the (geometric) mean noise power spectral
  density in the ALLEGRO and LIGO I detectors. The heavy solid line is
  the amplitude of the terrestrial noise, (mis)interpreted as a
  stochastic gravitational wave signal. Note how, after approximately
  3 months, the observations are no longer consistent with stochastic
  gravitational wave amplitude significantly less than the amplitude
  of the correlated terrestrial noise.
  The thin line marks the upper limit on the stochastic signal, again
  as a function of time, when the modulation technique described in
  this paper is used to make the measurement. The measurement is no
  longer biased by the terrestrial noise and the upper limit is less
  than the correlated terrestrial noise amplitude, in this example, in
  0.45~y.
  B) The integration time needed for the upper limit, estimated by the
  modulation technique described here, to be less than the amplitude
  of the correlated terrestrial noise amplitude (i.e., to reach the
  crossing point marked by the bold circle in panel A) as a function
  of the cross-spectral density. Curve (i) is for \emph{LIGO I +
    ALLEGRO}; (ii) is for \emph {advanced LIGO + Upgraded ALLEGRO} (cf. \ 
  table \ref{tbl:opChar}). }\label{figure:perf}
\end{figure}

Consider, for example, the current generation of ALLEGRO and LLO
detectors jointly observing in the presence of a correlated noise
\begin{equation}
    \left|S_{AL}(f_{0})\right| =
    10^{-4}\left(S_{A}S_{L}\right)^{1/2}. 
\end{equation}
Assume that the stochastic gravitational wave signal amplitude is much
smaller: $\Omega_{\text{GW}}$ equal to $10^{-9}$.  Suppose first that
we are ignore the possibility that the correlated noise component
(represented by $S_{AL}$) may be present. Then we would leave the
ALLEGRO detector orientation fixed in such a manner as to maximize the
overlap with LLO. The dashed lines in figure \ref{figure:perf}A show,
as a function of observing time, the 90\% confidence interval
(following the construction of \cite{feldman98a}) associated with an
observed cross-correlation $C$ (cf.\ eq.\ \ref{eqn:C-defn}) equal to
the ensemble mean $\bar{C}$.
After approximately 0.25~y this most likely observation is clearly no
longer consistent with the actual stochastic gravitational wave
background amplitude, owing to the systematic error made by excluding
the possibility of a correlated noise background. As the observation
time increases, the confidence interval on $\Omega_{\text{GW}}$
shrinks, asymptoting on the amplitude of the correlated noise
($S_{AL}$) interpreted as a stochastic gravitational signal.

On the other hand, suppose we admit the possibility of a correlated
noise background, of unknown cross-spectral density, changing the
orientation of the ALLEGRO detector mid-way through the observation in
order that we can construct $C_0$ (cf.\ eq.\ \ref{eqn:C0-defn}), which
is independent of $S_{AL}$. Again referring to figure
\ref{figure:perf}A, the thin gray line shows the 90\% confidence
interval (following the construction of \cite{feldman98a}) on
$\Omega_{\text{GW}}$ when the observed $C_0$ is equal to its ensemble
mean. The confidence interval is, in this case, \emph{always}
consistent with a stochastic gravitational wave background amplitude
$\Omega_{\text{GW}}$ of $10^{-9}$.  Additionally, in less than 0.45~y
this 90\% bound limits the signal amplitude to less than the
correlated background noise amplitude.

In this example the modulation technique described here provides, in
approximately 0.45~y, a bound on the stochastic signal below the
correlated noise background amplitude, interpreted as a stochastic
gravitational wave signal.
Figure \ref{figure:perf}B shows the integration period required, using
this technique, to limit the stochastic background to an amplitude
less than the correlated noise background as a function
$|S_{AL}(f_{0})|^{1/2}$.  The solid line corresponds to the \emph{LIGO
  I/ALLEGRO} detector pair while the dashed line, labeled (ii),
corresponds to the \emph{Advanced LIGO/Upgraded ALLEGRO} detector pair.
Since, with fixed detectors, the upper limit on the stochastic signal
strength is always above the amplitude of the correlated noise, figure
\ref{figure:perf}B shows that, after 1~y of observation with
\emph{LIGO I + ALLEGRO}, an unaccounted for correlation in the
background at the level of $\sqrt{S_{AL}(f)}\approx 3\times10^{-23}
~1/\sqrt{Hz}$ compromises the measurement. Similarly, after 1~y of
observation with \emph{Advanced LIGO + Upgraded ALLEGRO}, a correlated
background with a strain spectral density of $\sqrt{S_{AL}(f)}\approx
2\times10^{-25}~1/\sqrt{Hz}$ will compromise a simple correlation
measurement that does not account properly for environmental
correlations.

\section {Discussion}
\label{section:dd}

Weber-bar gravitational wave detectors, like ALLEGRO, have placed
progressively lower upper limits on the stochastic gravitational wave
background amplitude near 900~Hz
\cite{hough75a,zimmerman80b,compton94a,astone99b,astone99c,astone00a}.
The present best upper limit on $\Omega_{\text{GW}}$, set by Astone
\emph{et al.}  at 900 Hz \cite{astone99b,astone00a}, used two
cryogenic bar detectors operating over a fairly long (600 km)
baseline. During a relatively short measurement time they saw no
excess correlation between the two bar signals. The planned
LIGO-Allegro observations will be made over a much shorter baseline
(40~Km); consequently, it correlated environmental noises may be more
significant and greater attention will need to be paid to identifying
and dismissing them.

In the presence of an undetected correlated background, the best upper
limit that can be set will eventually become limited by the bias
introduced by the presence of unaccounted correlated detector noise.
For the ALLEGRO-LLO pair, correlated noise at a level of $\approx
10^{-4}$ the geometric mean noise spectral density of the two
detectors will compromise the observation after less than 1 year of
integration time. The possibility that geophysical or other
terrestrial correlations could compromise the observation in a single
orientation of the detector pair are discussed
in\cite{christensen92a,flanagan93a,allen99b}; however, those
discussions focus on estimating the maximum allowable correlated
background for determining a given value for an upper limit.  We show
here how rotating the ALLEGRO detector allows us to distinguish
between a correlated background and a weak, stationary stochastic
gravitational wave signal, removing this limit on the sensitivity of a
measurement.

The ALLEGRO detector has recently been moved to new quarters and
mounted on an air-bearing \cite{allegroII}, allowing its orientation
with respect to LLO be modulated in manner described here.  In a real
observation campaign, the frequency with which the detector can be
re-oriented is limited from above by the loss of observing time that
is introduced through the disturbance of rotating of the detector and
the need for the detector to settle down after rotation.  Similarly,
it is limited from below by the desire to make measurements in
multiple orientations.  A reasonable compromise would be to re-orient
ALLEGRO every $\approx 3 - 5$ months, while making sure that that
exact period is \emph{not} commensurate with any obvious seasonal or
annual cycles.

The degree of improvement possible with this technique will depend on
(i) whether the terrestrial background has any component with a
characteristic quadrupolar signature that aliases into the stochastic
gravitational wave background signature and (ii) the degree to which
the background is stationary over the separate periods of measurement
in different orientations.  Nonetheless, an experiment that modulates
the stochastic gravitational wave background signature will improve
the quality of any long term observation of $\Omega_{\text{GW}}$.

\acknowledgments 

The authors thank both Benoit Mours and Andrea Vicere' for helpful
suggestions and comments during writing of this paper. We are also
indebted to Rai Weiss for pointing out to us that the idea of signal
modulation originated with Dicke in the eponymous radiometer that was
invented at the MIT Radiation Laboratory at the end of World War II.
We also acknowledge the assistance of Evan Mauceli, Warren Johnson and
William Hamilton in providing us with the details of the ALLEGRO
resonant bar performance.  Finally, we thank Ken Strain for providing
the advanced LIGO narrowband performance projection.

This work has been partially supported by NSF grants PHY98-00111 and
PHY99-96213.  LIGO Laboratory is supported by the NSF under
cooperative agreement PHY92-10038.

This document has been assigned LIGO Laboratory document number 
LIGO-P000012-A-E.





\appendix
\section {The optimal filter in the presence of 
correlated detector noise}
\label{app:Q}

In this appendix we evaluate the kernel $Q$ (cf.\ eq.\ 
\ref{eqn:C-defn}) that maximizes the signal-to-noise ratio $\rho^2$ in
equation \ref{eqn:defn-rho}, when a correlated noise component is
present.

The general two-detector cross-correlation is given by:
\begin{equation}
   C(T) := \int_{-T/2}^{T/2}dt'\int_{-T/2}^{T/2}dt'' s_{1}(t')
   s_{2}(t'') Q(t'-t'')
\end{equation}
where $s_{i}(t)$ are the signals from the two interferometers and
$Q(\tau)$ is the optimal filter kernel to be determined.  $s_{i}(t) =
h_{i}(t)+n_{i}(t)$, where $h_{i}(t)$ is the signal and $n_{i}(t)$ is
the environmental-plus-instrumental noise in each detector.

The correlation can also be expressed in the frequency domain.  With
the definition of the Fourier transform of a function $r(t)$ given by:
\begin{equation}
\tilde r(f):\equiv \int_{-\infty}^{\infty}dt\ 
e^{-i2 \pi f t}\ r(t),
\end{equation}
the correlation measurement is also equal to,
\begin{equation} 
C(T) := \int_{-\infty}^{\infty}df\ \int_{-\infty}^{\infty} df'\ 
\delta_T(f-f') \tilde s_1^*(f) \tilde s_2(f') \tilde Q(f')\ .
\label{eqn:Sfreq}
\end{equation}
Here $\tilde s_i(f)$, and $\tilde Q(f)$ are the Fourier transforms of
the signals $s_i(t)$ and $Q(t-t')$. For real $Q(t-t')$, the positive
and negative frequency values of its Fourier transform are related by
$\tilde Q(-f)=\tilde Q{}^*(f)$.  The function $\delta_T(f-f')$ is the
finite-time approximation to the Dirac delta function $\delta(f-f')$
defined by
\begin{equation} 
\delta_T(f):=\int_{-T/2}^{T/2}dt\ e^{-i2\pi ft}
:= {\sin(\pi f T)\over\pi f},
\label{eqn:delta_T(f)}
\end{equation}
which arises owing to the windowing property of the finite duration
($T$) measurement.

Generalizing on \cite{allen99b}, both the signals and the noise are
now assumed to be correlated:
\begin{mathletters}
\begin{eqnarray}
    \langle \tilde h^{*}_{1}(f)\tilde h_{2}(f')\rangle  &:=& \delta(f-f') 
    {3H_{0}^{2}\over20\pi^{2}|f|^{3}} \Omega_{\text{{GW}}}(|f|) 
    \gamma(|f|)\\
        \langle \tilde n^{*}_{i}(f) \tilde n_{j}(f')\rangle &:=&{1\over2}
        \delta(f-f')S_{ij}(|f|)\\
        \label{eqn:nnnn}\langle \tilde n^{*}_{i}(f) \tilde n_{j}(f') \tilde n^{*}_{i}(f'') 
        \tilde n_{j}(f''')\rangle &:=&{1\over4} \left( S_{i}(|f|)S_{j}(|f'|) 
        \delta(f+f'')\delta(f'+f''')
        +\right.\\
        \nonumber &&  S_{ij}(|f|)S_{ij}(|f''|) 
        \delta(f-f')\delta(f''-f''')+\\
        \nonumber && \left.S_{ij}(|f|)S_{ij}(|f'|) 
        \delta(f-f''')\delta(f'-f'') \right)
\end{eqnarray}
\end{mathletters}

Equation \ref{eqn:nnnn} follows from the moment theorem for real
Gaussian noise \cite[Chapter 3]{bendat86a}. $S_{i}(|f|)$ is the power
spectral density of the detector noise for the $i^{th}$ detector and
$S_{ij}(|f|)$ is the cross-spectral density for the pair $(i,j)$. The
factor of $1/2$ in the power spectral densities arises from the
definition of the power spectrum as a function for $f>0$.

The cross-correlated signal is a function of relative detector
orientation through the overlap reduction factor. Call $C_{+,-}$ the
integrated measurement made for two orientations so that the
contribution to each from the stochastic gravitational wave background
is given by
\begin{equation}
\langle h^{*}_{1}(f) h_{2}(f)\rangle  =
\pm {3H_{0}^{2}\over20\pi^{2}|f|^{3}} \Omega_{\text{{GW}}}(|f|) 
\gamma(|f|)
\end{equation}
For a total measurement interval T, the strategy will be to measure
first in one orientation for a fraction of the time, and then to
measure in the other orientation for the remainder of the time.
Consider the case in which the two intervals are each equal to $T/2$.
The two of measurements have the following expectation values:
\begin{equation}
\langle C_{+,-}\rangle:= T/2
\int_{-\infty}^{\infty} df\ \left(\pm {3 H_0^2 \over 20 \pi^2}
{\Omega_{\rm gw} (|f|)\over|f|^{3}} 
\gamma(|f|) + S_{12}(f)\right) Q(f).
\label{eqn:mu_final}
\end{equation}

The two quantities $C_{+,-}$ can be used to eliminate from the
expectation value of the derived signal the effect of the correlated
detector noise:
\begin{equation}
\langle C\rangle:=\langle C_{+} - C_{-}\rangle:= T\,
\int_{-\infty}^{\infty} df\ \left({3 H_0^2 \over 20 \pi^2}\,
{\Omega_{\rm gw} (|f|)\over|f|^{3}} 
\gamma(|f|)\right) Q(f)\, .
\label{eqn:mu_final2}
\end{equation}
The optimal filter function follows from maximizing the the
signal-to-noise ratio
of the cross-correlation:%
\begin{equation}
{\rm SNR}:={\langle C \rangle\over\sigma_{C}},
\label{eqn:SNR_display}
\end{equation}
where $\sigma_{C}^2 = \langle C^{2}\rangle -\langle C\rangle ^{2}$ is
the variance of the cross-correlation signal $C$. Now the measurements
$C_{+,-}$ take place during two distinct intervals each of duration
$T/2$. It is reasonable to assume that the fluctuations in these two
measurements are not correlated. Then,
\begin{mathletters}
  \begin{eqnarray}
    \sigma_{C}^2 &:=& \langle C^{2}\rangle -\langle C \rangle^{2}\\
    &:=& \langle (C_{+} - C_{-})^{2} \rangle - \langle C_{+} - C_{-} 
    \rangle^{2}\\
    &:=&2\sigma_{C_{+}}^{2}
  \end{eqnarray}
\end{mathletters}
The last simplification follows from the assumed stationarity of the
noise and equal measurement intervals with identical statistical
properties. 

The technique used to evaluate $\sigma_{C_{+}}^2$ is similar to that
of \cite[Eq. 3.61]{allen99b}.  First assume that the noise intrinsic
to the two detectors is much larger in magnitude than the stochastic
gravitational wave background.  Then the variance of the measurement
will be dominated by the detector noise and not the astrophysical
signal.  In this case,
\begin{mathletters}
    \begin{eqnarray}
\sigma_{C_{+}}^2
&:=&\langle C_{+}^2\rangle -\langle C_{+}\rangle^2\\
&\approx&
\int_{-\infty}^\infty df\ \int_{-\infty}^\infty df'
\int_{-\infty}^\infty dk\ \int_{-\infty}^\infty dk'
\delta_{T\over2}(f-f')\delta_{T\over2}(k-k')
\nonumber\\ &&
 \times \left( 
   \left<\tilde n_1^*(f)\tilde n_2(f')
     \tilde n_1^*(k)\tilde n_2(k')\right> 
   -
   \left<\tilde n_1^*(f)\tilde n_2(f')\right>
   \left< \tilde n_1^*(k)\tilde  n_2(k')\right> 
\right) 
 \tilde Q(f')\tilde Q(k')\\
&\approx&
\int_{-\infty}^\infty df\ \int_{-\infty}^\infty df'
\int_{-\infty}^\infty dk\ \int_{-\infty}^\infty dk'
\delta_{T\over2}(f-f')\delta_{T\over2}(k-k') \tilde Q(f')\tilde Q(k')
\nonumber\\
&& \times ~{1\over4}\left[ 
S_{1}(|f|)S_{1}(|f'|) \delta(f+k)\delta(f'+k') + 
S_{12}(|f|)S_{12}(|f'|)\delta(f-k')\delta(f'-k) 
\right],
\end{eqnarray}
\end{mathletters}
where $S_1$ and $S_2$ are the power spectral densities of the noise in
detectors $1$ and $2$, and $S_{12}$ is the cross-spectral density of
the noise in the two detectors.

The integrals over $k$, $k'$ collapse due to the $\delta$-functions,
leaving:
\begin{mathletters}
\begin{eqnarray}
\sigma_{C_{+}}^2
&\approx&{1 \over 4} \int_{-\infty}^\infty df\ \int_{-\infty}^\infty df'
\delta_{T\over2}^2(f-f') \left(S_{1}(|f|)S_{2}(|f'|) + 
S_{12}(|f|)S_{ij}(|f'|)\right) \tilde Q (f)\tilde Q^*(f')\\
&\approx&{T\over 8}\int_{-\infty}^\infty df\ \left(S_{1}(|f|)S_{2}(|f|) + 
S_{12}(|f|)^{2} \right) 
|\tilde Q (f)|^2.
\end{eqnarray}
\end{mathletters}
%
%
The last expression is obtained by approximating one of the
finite-time delta functions, $\delta_{T\over2}(f-f')$ , as an ordinary
Dirac delta function while evaluating the other at $f=f'$.

Thus the signal and its variance are given by:
\begin{mathletters}
\begin{eqnarray}
\ \langle C\rangle&:=&\langle C_{+} - C_{-}\rangle:= T
\int_{-\infty}^{\infty} df\ \left({3 H_0^2 \over 20 \pi^2}
|f|^{-3}\ \Omega_{\rm gw} (|f|)
\gamma(|f|)\right) Q(f),\\
\sigma_{C}^2
&\approx&{T\over 4}\int_{-\infty}^\infty df\ (S_{1}(|f|)S_{1}(|f|) + 
S_{12}(|f|)^{2}) 
|\tilde Q (f)|^2 .
\end{eqnarray}
\end{mathletters}

Derivation of the optimal function $\tilde Q(f)$ proceeds along the 
same lines as in \cite{allen99b}. The analogous result is:
\begin{equation} 
\tilde Q(f) \propto 
{\gamma(|f|) \Omega_{\rm gw}(|f|) \over |f|^3 \left( S_{1}(|f|)S_{1}(|f|) + 
S_{12}(|f|)^{2} \right) } ,
\label{eqn:optimal2}
\end{equation}
It can be seen that the effect of correlated inter-detector noise is
to replace the product of the single-detector power spectral densities
with a sum of this product and the square of the contribution coming
from the cross-spectrum. If the cross-spectral density is sufficiently
small compared to the single-detector noise spectral densities, then
the effect on the optimal filter will be correspondingly small.

\begin{table}
\squeezetable
  \caption{Geographic Data for LIGO Livingston Laboratory (LLO). 
    Positions are with respect to the Earth Centered Frame (ECF):
    $\hat{z}$ pierces the earth at the north pole, $\hat{x}$ pierces
    the earth at the intersection of the prime meridian and the
    equator, and $\hat{y} = \hat{z}\times\hat{x}$}
    \label{tbl:coordsL}
    \begin{tabular}{|c|c|c|c|}
    Quantity&Symbol&Value&Units\\
    \hline    
     \centering LLO Vertex&$\{X_{E},Y_{E},Z_{E}\}$& 
     $\{-74276.044,-5496283.721, 3224257.018\}$ & $m$\\
     &$\{h, \phi, \lambda \}$ & $\{-6.568,N30^{\circ}~33'~6.871",
     W90^{\circ}~48'~50.229"\}$&\{m,dms,dms\} \\
    \hline
    {$X$ arm} &&&\\ {unit vector}&
    $\{
    \widehat{\bbox{X}}\cdot\widehat{\bbox{n}}_{x},
    \widehat{\bbox{X}}\cdot\widehat{\bbox{n}}_{y},
    \widehat{\bbox{X}}\cdot\widehat{\bbox{n}}_{z}
    \}$&
    \{-0.954574,-0.1415805,-0.2621887\}& 
    ECF\\
\hline
    {$Y$ arm} &&&\\{ unit vector}&
    $\{
    \widehat{\bbox{Y}}\cdot\widehat{\bbox{n}}_{x},
    \widehat{\bbox{Y}}\cdot\widehat{\bbox{n}}_{y},
    \widehat{\bbox{Y}}\cdot\widehat{\bbox{n}}_{z}
    \}$&
    \{0.2977412,-0.4879104,-0.8205447\}& 
    ECF\\
\hline
     {\centering Bearing of} &&& {\centering Reference is}\\ 
     {ALLEGRO at} &&$S66.88^{\circ}W$ & {geodetic } \\ 
     {LLO Vertex}&&&north\\
     \hline
    {\centering Angle between} &&&\\ 
    {LLO arm}&&&Degrees, \\
 
 {bisector and}&$\sigma_{L}$&$39.59^{\circ}$ (Bearing: 
 $S27.28^{\circ}W$)&measured\\ 
    {LLO-}&&&$CCW$ from\\ 
    ALLEGRO&&&baseline\\
    {baseline}&&&\\
\end{tabular}
\end{table}
\begin{table}
  \caption{Geographic Data for ALLEGRO Bar Detector at LSU. For the 
    definition of the Earth Centered Frame see the caption of table 
    \ref{tbl:coordsL}}
  \label{tbl:coordsA}
  \begin{tabular}{|c|c|c|c|}
    Quantity&Symbol&Value&Units\\
    \hline    
    ALLEGRO Vertex&$\{X_{E},Y_{E},Z_{E}\}$&$\{-113258.848,5504077.706,
    3209892.353\}$ & $m$\\ 
    &$\{\phi,\lambda\}$ &$\{N30^{\circ}~24'~45.110",
    W91^{\circ}~10'~43.766"\}$& $\{dms,dms\}$ \\ 
    \hline
    Bearing of LLO &&&Reference is\\
    Vertex at &&$N66.67^{\circ}E$ & geodetic north\\
    ALLEGRO                              &&&\\
    \hline
    Angle between &&Correlation maximum:$\gamma_{\text{max}}(921 Hz) =
    0.953$&Degrees,\\
    ALLEGRO bar axis&&$-5.60^{\circ}$ (Bar axis bearing: 
    $S72.08^{\circ}W$)&measured\\
    and LLO-ALLEGRO &$\sigma_{A}$&Correlation null: $\gamma_{null}(921 Hz) = 
    0.0 $& $CCW$\\
    baseline for&&$40.52^{\circ}$ (Bar axis bearing: $S26.15^{\circ}W$)&
    from baseline\\ 
    various values&&Correlation minimum:$ \gamma_{\text{min}}(921 Hz)
    = -0.893$&\\ 
    of correlations&&$84.60^{\circ}$ (Bar axis bearing:
    $S17.92^{\circ}E$)&\\ 
    \hline
    LLO - ALLEGRO&&&\\
    baseline distance&$L$&\centering 42269.951&m\\
    \hline
    Angle subtended by &&&\\
    LLO - ALLEGRO &$\beta $&$0.358^{\circ}$  &Degrees\\
    baseline at center&&&\\
    of Earth&&&\\
  \end{tabular}
\end{table}
\newpage
    \begin{table}
    \caption{Operational characteristics of the LLO and ALLEGRO detectors.}
    \label{tbl:opChar}
    \begin{tabular}{|p{120pt}|p{120pt}|p{120pt}|c|}
    \centering Quantity&\centering Lower ALLEGRO resonant frequency, 
    $f_{<}$&\centering Upper ALLEGRO resonant frequency, $f_{>}$&Units\\
    \hline
    \hline
    \centering Frequency &  \centering$896.8$ & \centering 920.3 & Hz \\
    \hline
   \centering  ALLEGRO bandwidth,~$\Delta f$&\centering $1$&
    \centering $1$ &  Hz \\ 
    \hline
   \centering  Upgraded ALLEGRO bandwidth\cite{allegroII},~$\Delta
    f$&\centering $50$&  
   \centering $50$ &  Hz \\
    \hline
    \hline
    \centering ALLEGRO sensitivity, $h(f)$ & \centering $1.8 \times
    10^{-21}$ &     
    \centering $0.85 \times 10^{-21}$ & $1/\sqrt{Hz}$ \\
    \hline
    \centering LIGO I sensitivity\cite{LIGOE95001802}, $h(f)$ &
    \centering $1 \times 10^{-22}$ &     
    \centering $1 \times 10^{-22}$ & $1/\sqrt{Hz}$ \\
    \hline
    \centering Adv.\ LIGO narrowband sensitivity\cite{ligoII}, $h(f)$ &
    \centering $2 \times 10^{-24}$ &    
    \centering $2 \times 10^{-24}$ & $1/\sqrt{Hz}$ \\
    \hline
    \hline
     \centering $\Omega_{\text{min}}$ for LIGO~I + ALLEGRO after 1
    year &\multicolumn{2}{c}{$1\times 10^{-1}$} \vline &at 90\%
    confidence \\ 
    \hline
    \centering $\Omega_{\text{min}}$ for adv.\ LIGO + Upgraded ALLEGRO
    after 1 year &\multicolumn{2}{c}{$2.9 \times 10^{-4}$} \vline &at
    90\% confidence \\ 
\end{tabular}
\end{table}


\end{document}